# VIRTUAL LEARNING ENVIRONMENT IS PLEASURE OR PRESSURE

**Mr. Subaveerapandiyan A**
**Ms. Ammaji Rajitha**

**Abstract**

The primary aim of this study intends the perception of students towards online learning in the covid-19 pandemic period. The pandemic has changed the traditional concepts of the education system and broken the functions of the educational institutions. But, they give it an opportunity to change pedagogy. The research paper discussed the students' opinions on online learning/virtual classroom learning. This study applied a qualitative approach and prepared a systematic questionnaire for data collection. The researcher collected the data from 258 students from different places in India and also, the disproportionate sampling used for data collection. The research mainly focused on the student's perception, the comfort and discomfort of e-learning, using electronic devices for communication, the virtual learning is a pleasure or pressure to the students, the digital skills of the students and their active performance. The study revealed that over 50% of the students are having excellent knowledge of digital skills. The students are attending online classes through their personal computers or laptops and phones. The teachers are allowing the students to ask questions and clear the doubt of the students. The study found that the students are losing social interaction with teachers, friends and cannot access the library because of online classes. Finally, the students felt that online learning is a pressure instead of pleasure.

## Introduction

The pandemic has created revolutionary changes in human life. Suddenly, the education system/ Over 90% of the students are not attending school. The education institutions tried to continue the legacy of education to all student communities at remote learning. The total education systems engaged The various information communication technologies used to continue the academic environment in the students' life. The government of India has taken it as a challenge and implemented various free digital online learning and teaching platforms and conducted training programs by the expert group. These online platforms are providing the quality of education to many social groups of people in India. The students and teachers improved their online learning skills to break the worst pandemic situation in all educational domains. They used the internet and various communication devices to continue education on the online learning platforms. The traditional classroom learning system completely converted into digital learning and accessing electronic learning material through the internet all over the world. They effectively involved the students and teachers in the online learning mechanism. The government moral standards of students' knowledge through flexible timings, fixed schedules and quality of content.

**Objectives of the Study**

- ➢ To find out the student's perception and opinion on online learning education.
- ➢ To know the merits and demerits of e-learning platforms.
- ➢ To know the use of electronic devices by the students to access the online learning education.
- ➢ To investigate the virtual learning is a pleasure or pressure to the students
- ➢ To determine the resources used for writing and reading an assignment
- ➢ To know the digital literacy skills among the students

**Previous Studies**

Shawaqfeh et al. (2020) investigated a study of pharmacy student's distance online learning experimentally in the COVID-19. This study, carried through a structured questionnaire to find the performance of the students in online learning. Finally, the investigator concluded that online learning became very difficult because of the lack of computer literacy among the students. Therefore, the school should conduct training and orientation programs to address students' needs for improving computer skills.

Rafique et al. (2021) investigated a study of readiness for online learning of Pakistani LIS students in the pandemic. The data were carried through a structured questionnaire, and SSP software AMOS analyzed data to determine the readiness and self-sufficiency skills of the students. Finally, the investigator concluded that the female students' self-efficiency of online communication was very low compared with male students. Therefore, the Department of Library and Information science should take orientation and training programs to improve online communication self-efficiency for better academic performance.

Bączek et al. (2021) investigated a study of students' perception of online learning during the Covid-19 pandemic. The data was collected through a structured questionnaire to analyze the clinical, social, and technical skills in online education and face-to-face learning. Finally, the investigators concluded that online learning gives the advantage of staying home and accessing electronic information. Still, it is not less effective than face-to-face learning to increase skill.

Deka (2021) investigated the study of factors influencing U.G and P.G students' engagement in online learning during the Covid-2019 pandemic in India. The data was collected through a structured questionnaire to investigate the impact created by arranging online education for students. The study revealed that students were affected more and faced lots of problems because of online classes. However, the students' interest, course content, learning environment, course design and administrative support influenced more students to engage in online learning.

Olayemi et al. (2021) conducted a study on students' perception and readiness towards online learning during the Covid-19 pandemic in Nigeria to identify the skills and competence of online learning. They used a structured questionnaire for data collection. The study has revealed that most respondents expressed positive responses and claimed they were very conversant with online learning. However, the readiness among the respondents also became low because of the high cost of data. The investigator suggested that the Nigerian government invest more in communication technology and its infrastructure by collaborating with other telecommunication companies and implementing technology-based learning through higher education institutions.

**Research Method**

The study adopted a survey method, prepared a structured questionnaire and shared it with Indian Colleges (Government and Private) and University students. The questionnaire is divided into two parts. Part-1 is socio-demographic details of the students and comprises five questions, and Part-2 is digital literacy skills and online learning and comprises nine questions. For this study, the targeted population is college and university students around India. The questionnaire was shared with 500 students, and the filled questionnaire was 258 (51.6%) responded—disproportionate sampling used for data collection. The data was collected from May to June 2021 throughout the google form, and a link was shared to the student's WhatsApp group with the help of institute faculty members. A five-point Likert scale was used for this study: Strongly Agree, Agree, Neutral, Disagree, and Strongly Disagree. Data collected and analyzed using simple frequency and percentages, mean, standard deviation and ANOVA test is conducted. For data analysis, we use SPSS.

**Data Analysis**

**Table 1. Socio-Demographic Characteristics of Respondents**

| Variable | Frequency | Percentage |
|---|---|---|
| **Age** | | |
| 17-20 | 95 | 36.8 |
| 21-25 | 125 | 48.4 |
| 26-30 | 28 | 10.9 |
| 31-40 | 8 | 3.1 |
| Above 40 | 2 | 0.8 |
| **Gender** | | |
| Male | 125 | 48.4 |
| Female | 133 | 51.6 |
| **Course Stream** | | |
| Arts | 112 | 43.4 |
| Science | 102 | 39.6 |
| Commerce & Management | 13 | 5 |
| Engineering | 31 | 12 |
| **College/University** | | |
| Private College | 45 | 17.4 |
| Govt College | 34 | 13.2 |
| University | 179 | 69.4 |
| **Location** | | |
| Urban | 88 | 34.1 |
| Semi-Urban | 61 | 23.6 |
| Rural | 109 | 42.3 |

Above table 1 shows the demographics details of the respondents. Age-wise, the highest respondents 48.4% they were 21 to 25 years old, the least respondents are 0.8% above 40 years old; Gender wise more respondents were 51.6% female and male respondents are 48.4%; course stream-wise most respondents are from Arts 43.4% of respondents, followed by Science 39.6% and the lowest from Commerce and Management 5%; College and University wise maximum responders were 69.4% were the University, and lowest respondents were 13.2% from Government college; location savvy massive respondents were 42.3% from rural, 34.1% were urban, and 23.6% were semi-urban.

**Table 2. IT Skills**

| IT Skills | Frequency | Percentage |
|---|---|---|
| Low | 31 | 12 |
| Medium | 189 | 73.3 |
| High | 38 | 14.7 |

The above table 2 reveals the IT Skills of students. A considerable part of the 73.3% of respondents was medium in IT skills, less than 14.7% of respondents only high and the least respondents 12% were only deficient in IT skills.

**Table 3. Digital Literacy Skills of the Students**

| Digital Literacy Skills | Very Good | Good | Acceptable | Poor | Very Poor |
|---|---|---|---|---|---|
| Typing | 67 (26%) | 129 (50%) | 53 (20.5%) | 9 (3.5%) | 0 (0%) |
| Web Search | 82 (31.8%) | 135 (52.3%) | 38 (14.7%) | 3 (1.2%) | 0 (0%) |
| Computer Literacy | 75 (29.1%) | 112 (43.4%) | 58 (22.5%) | 13 (5%) | 0 (0%) |
| Internet Literacy | 77 (29.9%) | 125 (48.4%) | 50 (19.4%) | 6 (2.3%) | 0 (0%) |
| Digital Literacy | 55 (21.3%) | 120 (46.5%) | 72 (27.9%) | 10 (3.9%) | 1 (0.4%) |
| MS Office | 78 (30.2%) | 115 (44.6%) | 53 (20.5%) | 10 (3.9%) | 2 (0.8%) |

Above table 3 presents students' self-rating of digital literacy skills. For example, 50% the half of the respondents were good at typing skills, 52.3% more than half of respondents were good in web search skills, 43.4% of the most of the respondents were good in computer literacy, 48.4% the highest respondents were good in Internet literacy skills, 46.5% most of the respondents were good in digital literacy skills, 44.6% the highest respondents were good in MS Office.

**Table 4. Way of accessing online course content**

| Way of accessing online course content | Frequency | Percentage |
|---|---|---|
| I have my own personal computer, laptop or tablet | 127 | 49.2 |
| I share a home computer, laptop or tablet | 14 | 5.4 |
| I borrowed a computer, laptop or tablet from someone outside of my home | 4 | 1.6 |
| I can only use my cell phone to access content | 113 | 43.8 |

Above table 4 shows that students are accessing online course content. 49.2% of students have their personal computer, laptop, or tablet, and 43.8% of respondents only have a cell phone to access course content, 5.4% of respondents share their home computer, laptop, or tablet, 1.6% of respondents borrow digital devices for accessing course content.

| Table 5. Digital information resources used for reading and writing an assignment | | |
|---|---|---|
| **Digital information resources used for reading and writing an assignment** | **Frequency** | **Percentage** |
| E-Books | 102 | 39.5 |
| E-Journals | 27 | 10.5 |
| E-Databases | 19 | 7.4 |
| E-Newspapers | 10 | 3.9 |
| Search Engines | 100 | 38.7 |

Table 5 represents digital information resources used for reading and writing an assignment. 39.5% most of the respondents used e-books, followed by 38.7% used search engines, 10.5% of respondents used e-journals and very few students used e-databases 7.4%, e-newspapers used 7.4%.

| Table 6. Student Opinion about Online Classes | | | | | | | |
|---|---|---|---|---|---|---|---|
| **Online Class and Student Opinion** | **Strongly Agree** | **Agree** | **Neutral** | **Disagree** | **Strongly Disagree** | **Mean** | **SD** |
| I have a positive impact on my studies due to online class | 30 (11.6%) | 54 (21%) | 91 (35.3%) | 46 (17.8%) | 37 (14.3%) | 2.97 | 1.19 |
| Online classes have increased my technological literacy | 41 (15.9%) | 108 (41.9%) | 78 (30.2%) | 17 (6.6%) | 14 (5.4%) | 3.56 | 1.01 |
| I feel online classes help me to gain more knowledge | 25 (9.7%) | 53 (20.6%) | 86 (33.3%) | 56 (21.7%) | 38 (14.7%) | 2.88 | 1.17 |
| I feel comfortable using online learning tools | 31 (12%) | 75 (29.1%) | 74 (28.7%) | 43 (16.7%) | 35 (13.5%) | 3.09 | 1.21 |
| I receive enough support and resources from my teacher | 51 (19.8%) | 111 (43%) | 61 (23.6%) | 20 (7.8%) | 15 (5.8%) | 3.63 | 1.06 |
| My teacher encourages discussion in an online class | 68 (26.4%) | 109 (42.2%) | 54 (21%) | 14 (5.4%) | 13 (5%) | 3.79 | 1.05 |

Scale Used: 1=Strongly Disagree, 2 Disagree, 3 Neutral, 4 Agree, 5 Strongly Agree *SD-Standard Deviation

Table 6 depicts the student's opinions about online classes. 35.3% most of the respondents were neutral, 21% were agreed, 17.8% have disagreed, 14.3% were strongly disagreed and 11.6% were strongly agreed; Online classes have increased students technological literacy 41.9% the highest respondents were agreed, 30.2% were neutral, 15.9% were strongly agreed, 6.6% have disagreed and 5.4% were strongly disagreed; Students free online classes help them to gain more knowledge 33.3% most of the respondents were neutral, 21.7% have disagreed, 20.6% were agreed, 14.7% were strongly disagreed and 9.7% were strongly agreed; students feel comfortable about using online learning tools 29.1% the majority of the respondents were agreed, 28.7% were neutral, 16.7% have disagreed, 13.5% were strongly disagreed and 12% were strongly agreed; Students receive enough support and resources from their teacher 43% the most of the respondents were agreed, 23.6% were neutral, 19.8% were strongly agreed, 7.8%online classes 42.2% the highest number of respondents were agreed, 26.4% were strongly agreed, 21% were neutral, 5.4% disagreed, and 5% were strongly disagreed.

**Table 7. Advantages and Disadvantages of Online Classes**

| Advantages and Disadvantages of Online Classes | Strongly Agree | Agree | Neutral | Disagree | Strongly Disagree | Mean | SD |
|---|---|---|---|---|---|---|---|
| The student-teacher interaction during online teaching & learning is comfortable | 24 (9.3%) | 71 (27.5%) | 75 (29.1%) | 48 (18.6%) | 40 (15.5%) | 2.96 | 1.2 |
| Professor allows to ask questions and clear doubts during online lectures | 81 (31.4%) | 114 (44.2%) | 48 (18.6%) | 6 (2.3%) | 9 (3.5%) | 3.97 | 0.95 |
| Home environment is suitable for participating online lectures | 32 (12.4%) | 69 (26.7%) | 64 (24.8%) | 49 (19%) | 44 (17.1%) | 2.98 | 1.28 |
| Accessing to online material is easy | 40 (15.5%) | 83 (32.2%) | 44 (24.8%) | 51 (19.8%) | 20 (7.7%) | 3.27 | 1.17 |
| Online class saves the food and accommodation cost | 91 (35.3%) | 93 (36%) | 44 (17.1%) | 20 (7.7%) | 10 (3.9%) | 3.91 | 1.08 |
| Reduced academic pressure | 36 (14%) | 69 (26.7%) | 56 (21.7%) | 53 (20.5%) | 44 (17.1%) | 3 | 1.31 |
| Technical problems | 72 (27.9%) | 97 (37.6%) | 55 (21.3%) | 19 (7.4%) | 15 (5.8%) | 3.74 | 1.11 |
| Lack of self-discipline | 57 (22.1%) | 79 (30.6%) | 77 (29.9%) | 26 (10%) | 19 (7.4%) | 3.5 | 1.15 |
| Lack of social interaction | 98 (38%) | 88 (34.1%) | 37 (14.3%) | 18 (7%) | 17 (6.6%) | 3.89 | 1.17 |
| Inability to focus on screens long-time | 128 (49.6%) | 70 (27.1%) | 28 (10.9%) | 15 (5.8%) | 17 (6.6%) | 4.07 | 1.19 |

Scale Used: 1=Strongly Disagree, 2 Disagree, 3 Neutral, 4 Agree, 5 Strongly Agree *SD-Standard Deviation

Above table 7 discussed online classes' advantages and disadvantages; the first six questions dealt with advantages, and the following 4 five questions dealt with disadvantages. The student's and teacher's interaction during online teaching and learning is comfortable 27.5% were agreed, professor allows to ask questions and clear doubts during online lectures 44.2% the most of the respondents were agreed and 31.4% were strongly agreed; the home environment is suitable for participating online classes 26.7% the most f the respondents were agreed and similarly, 24.8% were neutral; accessing of online study materials are easy 32.2% most of the respondents were agreed; online classes save the food and accommodation cost 36% the most of the respondents were agreed and similarly 35.3% were strongly agreed; reduced academic pressure 26.7% the most of the respondents were agreed; students faced while attending an online class in technical problems such as connectivity issue, data issue and security issue and so on the most of the respondents 37.6% were agreed and followed by 27.9% were strongly agreed, lack of self-discipline 30.6% the majority of the respondents were agreed and similarly 29.9% neutrals well 22.1% strongly agreed; lack of social interaction 38% the majority of the respondents were strongly agreed followed by 34.1% were agreed and 14.3% were neutral; inability to focusing on screen long time 49.6% almost half of the respondents were strongly agreed and 27.1% were agreed.

**Table 8. Elements of regular education missing from online education**

| Regular education missing from online education | Very much | Much | Quite | A Little | Not at all |
|---|---|---|---|---|---|
| Social interaction | 152 (58.9%) | 54 (21%) | 24 (9.3%) | 15 (5.8%) | 13 (5%) |
| Fellow students | 145 (56.2%) | 59 (22.9%) | 30 (11.6%) | 13 (5%) | 11 (4.3%) |
| Teachers | 136 (52.7%) | 63 (24.4%) | 33 (12.8%) | 15 (5.8%) | 11 (4.3%) |
| Library | 144 (55.8%) | 50 (19.4%) | 28 (10.9%) | 10 (3.9%) | 26 (10%) |
| Classrooms | 164 (63.6%) | 46 (17.8%) | 19 (7.4%) | 15 (5.8%) | 14 (5.4%) |

Above table 8 and below figure 1 elements of regular education are missing in online classes. More than half the respondents were missing very much in social interaction 58.9%, fellow students 56.2%, teachers 52.7%, library 55.8% and classrooms 63.6%.

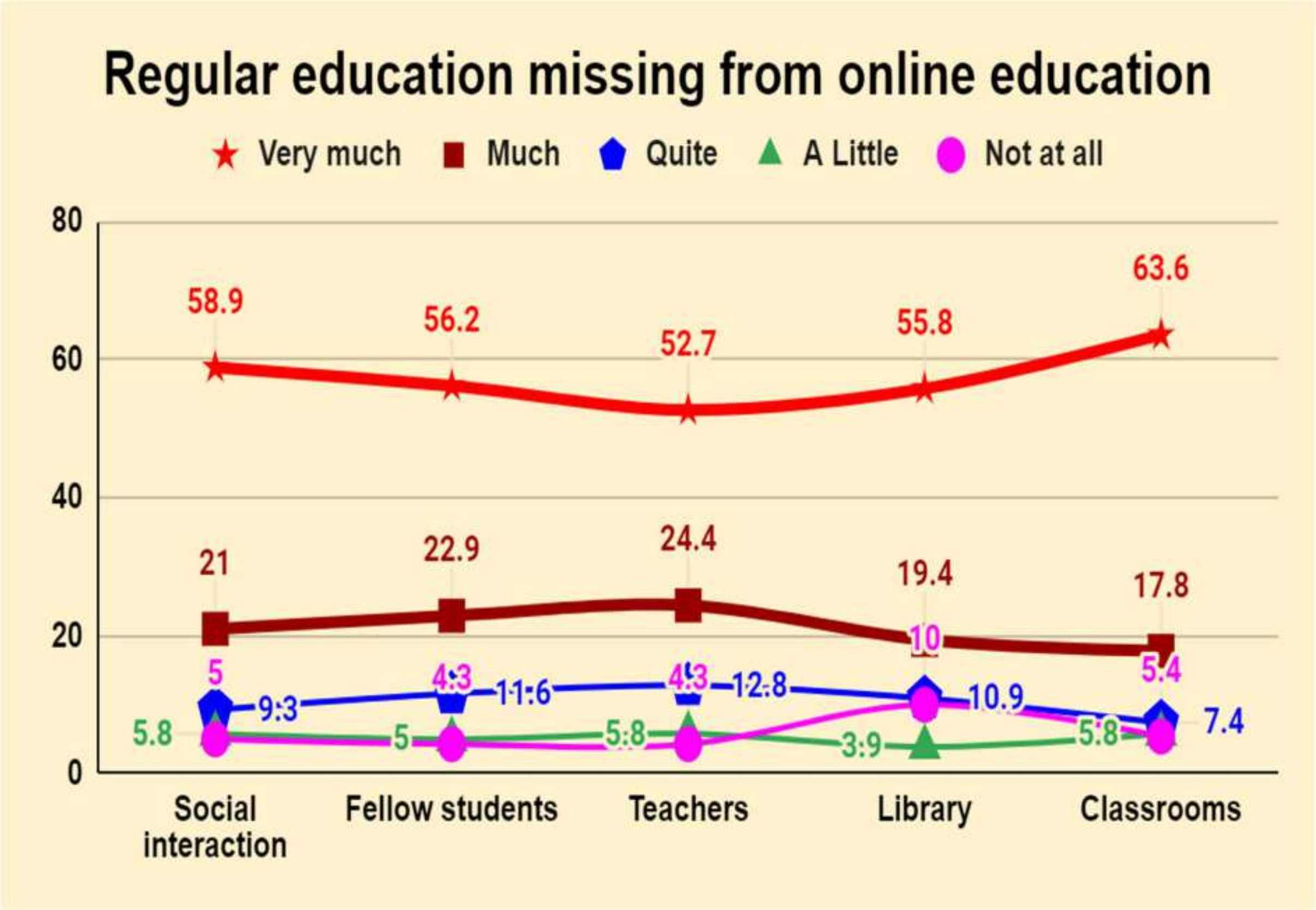

**Figure 1. Regular education missing from online education**

| Table 9. Online Learning Vs Classroom Learning | | | |
|---|---|---|---|
| **Online Learning Vs Classroom Learning** | **Online** | **Offline** | **Both** |
| Interaction among students and teacher is better in | 20 (7.8%) | 207 (80.2%) | 31 (12%) |
| We can get access to a good amount of study material in | 40 (15.5%) | 150 (58.1%) | 68 (26.4%) |
| Coverage of content of the particular topic is more in | 39 (15.1%) | 176 (68.2%) | 43 (16.7%) |
| Clarification of doubts is easier in | 28 (10.9%) | 193 (74.8%) | 37 (14.3%) |
| Appearing for internal assessments is easier in | 80 (31%) | 135 (52.3%) | 43 (16.7%) |
| Mode of learning you prefer during the pandemic | 184 (71.3%) | 47 (18.2%) | 27 (10.5%) |
| Mode of learning you prefer after the pandemic | 30 (11.6%) | 194 (75.2%) | 34 (13.2%) |

Table 9 shows which is better and easier online learning vs classroom learning. Interaction among students and teachers are better in 80.2% of respondents were feel offline, students can get a good amount of study materials in 58.1 respondents answered offline, coverage of the content of the particular topic is more in 68.2% were responded offline, clarification of doubts easier in 74.8% respondents emphasized offline, appearing for internal assessment is easier in 52.3% felt offline is comfortable, mode of learning during the pandemic suitable is 71.3% online and after pandemic the expect 75.2% were offline.

| **Table 10. Opinion on online learning is** | | |
|---|---|---|
| **Opinion on online learning is** | **Frequency** | **Percentage** |
| Pleasure | 25 | 9.7 |
| Pressure | 83 | 32.2 |
| Both | 150 | 58.1 |

Above table 10 and below figure 2 presents the responses about online learning as pleasure or pressure. 58.1% of the respondents feel online learning is a pleasure and pressure, 32.2% of respondents emphasized its pressure, and at least 9.7% of respondents felt the pleasure.

**Figure 2. Opinion on online learning**

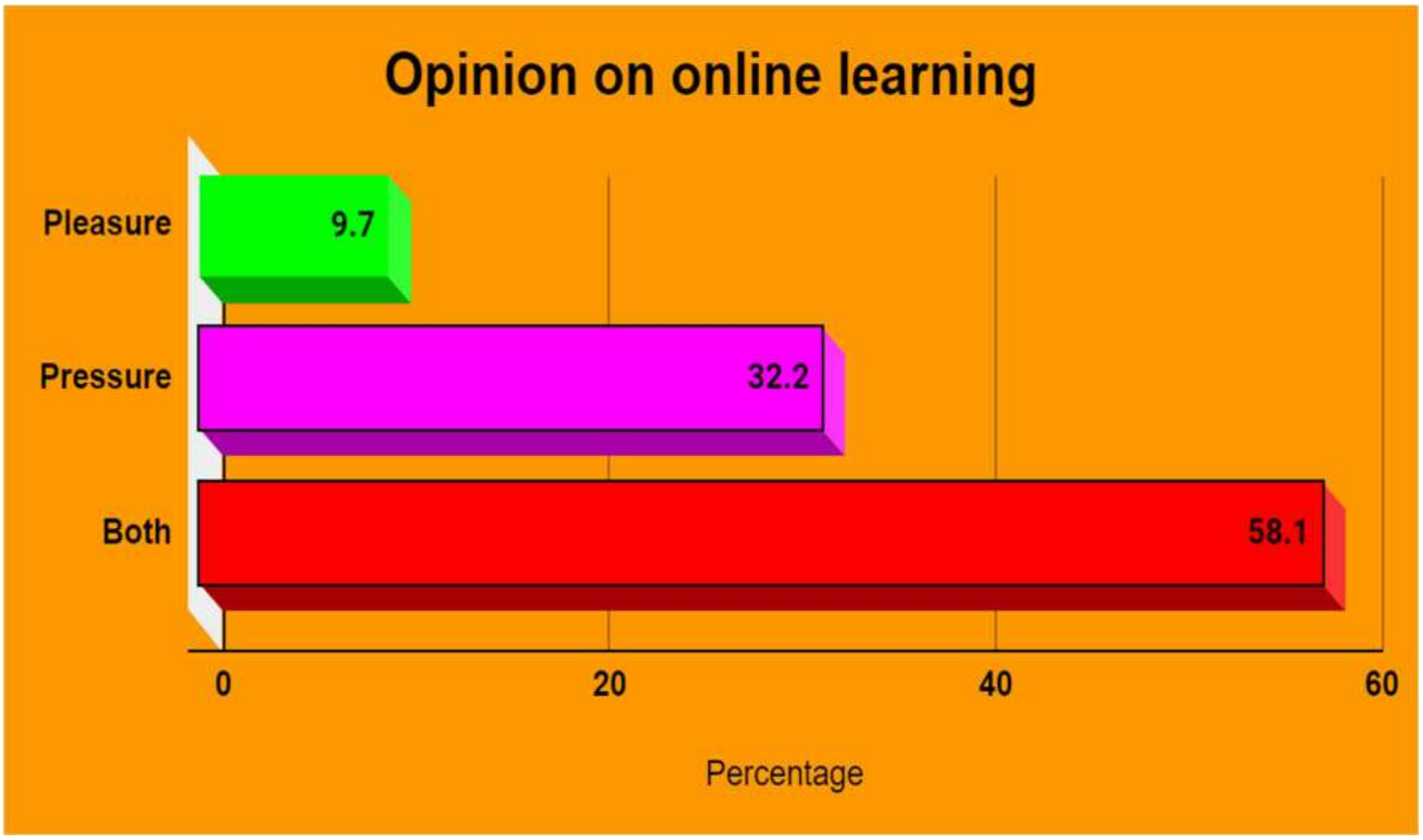

**Findings of the Study**

- It enlightens that 48.4% of the respondents are between 21-25 old.
- It determined that 73% of the respondents were having medium skills in IT.
- It showed that 50% of the respondents were very good at typing.
- It enlightens that 52% of the respondents were good in web search skills.
- It is shown that 46% of the majority of respondents were good in literary skills.
- It found that 49.2% of themaccessed their online course content by using their personal computers, laptops and tablets.
- The study result showed that most of the respondents mean 39.5% of respondents were using E-Books to access digital information resources for their reading and writing assignments.
- The study showed that 35.3% of neutral opinionswere there on the advantages and disadvantages of online classes.
- It found that 29.5% of respondents showed a neutral opinion on being comfortable with online teaching, learning and also the interaction between the students and teacher.
- The study showed that 31.4% of respondents said that the professors allowasking questions and clear doubts during online lectures.
- It found that 26.7% of respondents agreed that the home environment is suitable for participating in the online lectures and also, 36% of respondents agreed online classes save the food and accommodation cost too.
- The study found that 26.7%, 37.6% and 30.6% of respondents agreed online classes reduced the academic pressure, technical problems and created a lack of discipline among the respondents
- The study showed that The online classes created and built, the lack of social interaction and inability to focus on screen long hours among the 38% and 49.6% of respondents.
- The study highlights that 58.9% of respondents missed social interaction because of online education.
- It found that 56.2%, 52.7%, 55.8%, 63.6% of respondents were missing fellow students, teachers, library and classrooms because of online classes.
- The study found the 80.2% of students felt the offline classroom learning is better than online learning
- It showed that 58.1% of respondents get access to a good amount of study material through classroom learning compared to offline classroom learning.
- It highlights that 68.2% and 74.8% of respondents felt that the coverage of the particular topic is more and the possibility to clarify doubts are easier in offline classroom learning only.
- It found that 52.3% of respondents said that appearing for internet assessment is easier in offline classroom learning only.

- The study found that 71.3% of respondents preferred the online mode of learning because of the COVID-19 pandemic and 75.2% of respondents preferred offline classroom learning offline.
- The study finally showed that 9.7% of respondents felt that online learning is a pleasure, but 32.2% of respondents felt that online learning is a pressure.

**Conclusion**

The Covid-19 pandemic created new learning ways in pedagogy. It has shown a more impact on education. The latest technology is used in teaching and learning mechanisms to give the quality of education to the students. In developing countries like India, most of the Indian students belong to different economic and social backgrounds, therefore the facilities which are available to attend the online classes are low. The students are accessing and attending their online classes by using their personal communication devices like computers or laptops, Phones etc. The students are facing lots of problems while attending the online classes; they are exposed to network connectivity issues and electrical power issues. The online classes showed more impact on respondents' education and economic conditions too. The respondents said that online classes reduced the cost of learning. The pandemic once again remembered for all the importance of offline classroom learning mechanisms. Physical classroom teaching is more effective and comfortable to communicate with teachers and the students can have good eye contact with teachers to be a part of classroom conversations and discussions. This study showed that the respondents are uncomfortable with online learning, they felt that online class is more pressure instead of pleasure. The students are willing to attend the offline classroom learning.

**Subaveerapandiyan A**

**Professional Assistant (Library)**
**Regional Institute of Education, Mysore, India**
**E-mail: subaveerapandiyan@gmail.com**
**&**
**Ammaji Rajitha**
**PhD. Research Scholar**
**Department of Library and Information Science**
**Central University of Tamil Nadu**
**E-mail: ammajirajitha810@gmail.com**